\documentclass[reprint,amsmath,aps,longbibliography]{revtex4-1}
\usepackage{graphicx,epstopdf,dcolumn,bm,mathtools,siunitx}
\begin{document}
\preprint{APS/123-QED}
\title{Thermoshape effect for energy harvesting with nanostructures}
\author{Altug Sisman$^{1}$}
 \email{Corresponding author, altug.sisman@physics.uu.se}
\author{Alhun Aydin$^{1,2,3}$}
\author{Jonas Fransson$^{1}$}
\affiliation{$^{1}$Department of Physics and Astronomy, Uppsala University, Box 516, 75120, Uppsala, Sweden \\
$^{2}$Nano Energy Research Group, Energy Institute, Istanbul Technical University, 34469, Istanbul, Turkey \\
$^{3}$Fritz Haber Research Center, Institute of Chemistry, The Hebrew University of Jerusalem, 91904, Jerusalem, Israel}
\date{\today}
\begin{abstract}
We propose a mechanism for nanoscale energy conversion, an electric voltage induced by a temperature gradient in a junction composed of the same material having exactly the same geometric sizes, but distinct shapes. The proposed effect appears as a result of only temperature and shape difference, hence it is called thermoshape effect. For GaAs quantum confined semiconductor nanostructures, we first introduce the existence of quantum shape effects on thermoelectric transport coefficients at ballistic regime. We show that the shape alone enters as a control parameter on transport properties of confined nanostructures. The thermoshape voltage is then calculated by using the Landauer formalism. Our calculations show that the thermoshape voltage has a constant value in the order of mV/K for the variation of chemical potential in non-degenerate regime and it decreases rapidly after entering weakly degenerate regime where it oscillates around zero within plus/minus 10$\mu$V/K magnitude. A persistent voltage range may pave the way for easier experimental demonstration of the effect. Our work explicitly shows how important the effect of overall geometry is in nanoscale thermoelectric materials, and can be utilized even if all sizes are the same. A thermoshape junction not only represents a viable setup for the macroscopic manifestation of quantum shape effects, but also constitutes their first possible device application.
\end{abstract}
\maketitle
\section{Introduction}
Thermoelectric devices utilize the electric current induced by a temperature difference and vice versa. Most of the thermoelectric literature has been focused on the enhancement of the efficiency of thermoelectric devices by materials design \cite{te3,tebook2013}. Beginning with the works of Hicks and Dresselhaus in 1993 \cite{te1,te2}, lower-dimensional, nanoscale thermoelectrics has attracted an extensive attention \cite{te3,tejunc2011a,tebook2013,tebook2014}. It has been shown that efficiency can be enhanced by density of states engineering such that sharper density of states causes energy filtering and contributes to the increments especially in Seebeck coefficient \cite{te3,te4,tebook2013,tebook2014}. This feature has opened the possibility of tailoring geometry and quantum size effects to design efficient thermoelectric devices \cite{mitchen,geo1,geo2,geo3,geo4,geo5,qsetransporteir,tenw,doi:10.1021/acs.jpclett.7b00507}.

While material design based on quantum size effects becomes a widely popular route in thermoelectrics, yet another approach is to take advantage from size effects by making a junction of the same materials but having different sizes. In thermoelectric devices, most often dissimilar materials are used as junctions. However, electrochemical potential difference can also be induced even when the same materials but having different sizes are made junction under an applied temperature gradient \cite{tse1}. This potential difference emerges from the distinct Seebeck coefficients of the junction materials due to classical and/or quantum size effects. This phenomenon, so called thermosize effect, is first proposed in 2004 and numerous studies have been done to investigate thermosize effects after since \cite{tsef2008a,tsef2008b,tsef2010a,tse2,tse3,tsef2011a,tsef2012a,tsef2012b,tse4,tsef2013a,tsef2013b,tsef2014a,tse5,tsef2018a,tsef2018b,tsegrap}.

\begin{figure}[t]
\centering
\includegraphics[width=0.48\textwidth]{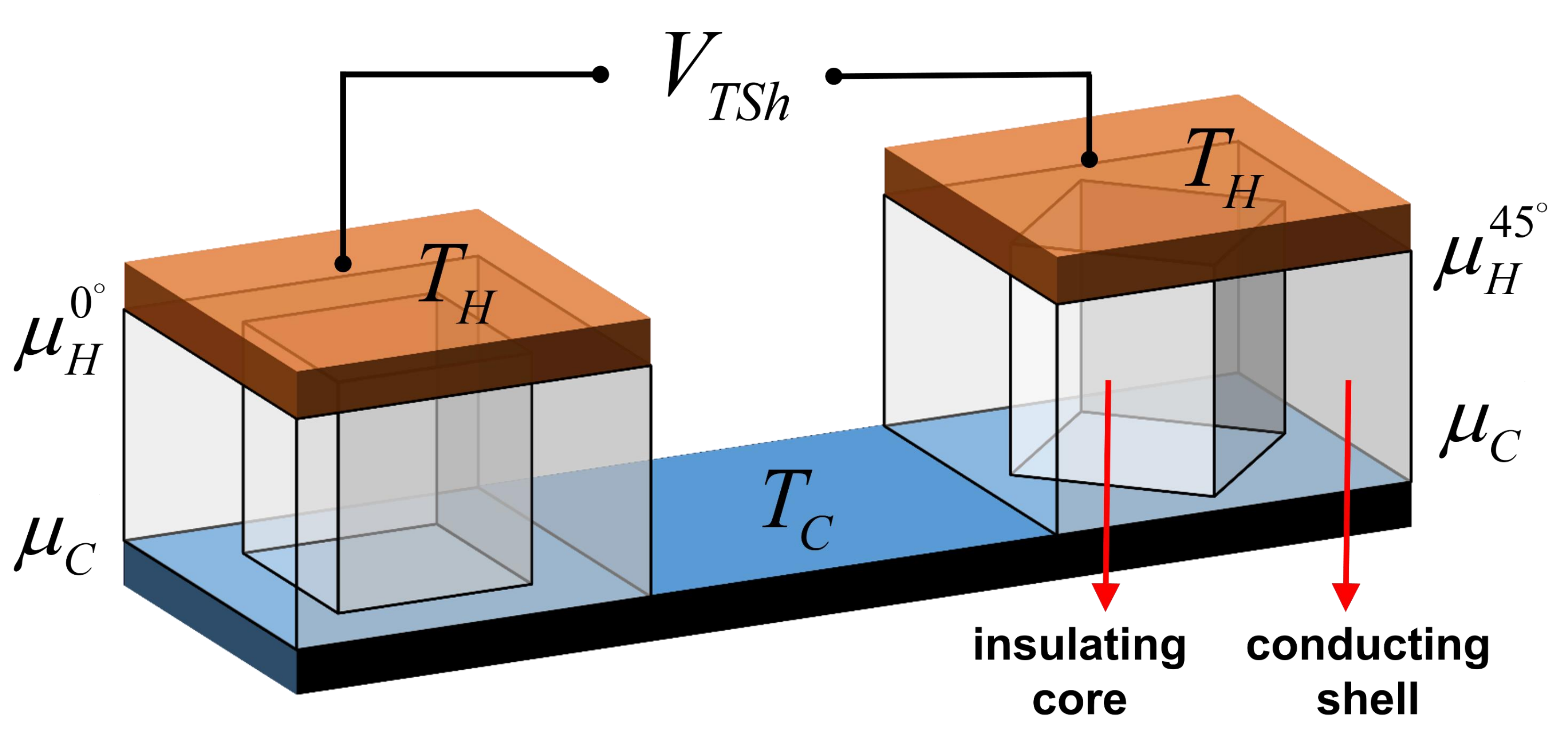}
\caption{A schematic of a thermoshape junction of core-shell (insulating-conducting) nanostructures having different confinement shapes in each pillar. The shape difference is associated with the configuration angle of the core structure, that are $\theta_L=0^{\circ}$ and $\theta_R=45^{\circ}$ for the left and right configurations of the junction components respectively. Thermoshape voltage $V_{TSh}$ is induced in such a junction because of temperature and shape differences.}
\label{fig:pic1}
\end{figure}

The sizes of a junction material are characterized by the so called geometric size variables, \textit{i.e.} volume $V$, surface area $A$, peripheral lengths $P$ and number of vertices $N_V$ (edges, point-like impurities or discontinuities on the boundaries in general). Note that geometric size variables include not only the volume, but also the lower-dimensional size elements. Then an intriguing question comes in; what if we make junction components out of the same materials even with the same sizes, but different purely in their shapes. Can we still get a reasonable electrochemical potential difference out of such a junction under a temperature difference? This question could not have been addressed so far, because until very recently there was no any clear mechanism that allows one to focus on purely shape effects by perfectly distinguishing them from the size effects. Even though size and shape effects typically interfere with each other, recently they have been completely separated by a size-invariant shape transformation and a new physical phenomenon called quantum shape effects has been introduced to control the material properties at nanoscale \cite{dsey}. Isolation of quantum size and shape effects is important, because they have some distinct influences on the physical properties of a confined system. As it's explored in Ref. \cite{dsey} in detail, size and shape effects can have quite different consequences on the system's thermodynamic properties because their underlying physical mechanism is different. Therefore, it is required to make an independent examination of how shape alone affects the transport properties of a confined structure. 

Motivated by the novel behaviors appearing in nanoscale thermodynamics due to quantum shape effects, here we apply this idea into nanoscale thermoelectrics, by creating a so called thermoshape junction, in which, unlike thermosize junctions, the difference between junction materials are not coming from their sizes but only from their shapes. A schematic view of a thermoshape junction composed of core-shell nanostructures is given in Fig. 1. The thermoshape junction is connected to a set of hot thermal reservoirs at temperature $T_H$ on the top and a cold thermal reservoir at temperature $T_C$ on the bottom end. At the cold end of the junction, the nanostructures are interconnected to each other both electrically and thermally, having chemical potential $\mu_C$. The junction nanostructures are made by the same material having the same sizes, \textit{i.e.} their $V$, $A$, $P$ and $N_V$ quantities are exactly the same. Size-invariant shape transformation is provided by the preparation of the non-conductive core structures differing in angular configuration by a certain degree $\theta$ (e.g. $\theta_L=0^{\circ}$ and $\theta_R=45^{\circ}$ in the Fig. 1) defined in transverse plane. Just like quantum size effects leading to the thermosize effect, we expect quantum shape effects to lead the thermoshape effect so that when temperature difference is maintained at both ends of a thermoshape junction, we expect an electric voltage at the hot end.

\section{The model and results}
Our physical model consists of a thermoshape junction under a temperature difference, Fig. 1, and we use quantum transport framework to investigate its properties. We first start by examining the thermoelectric properties of each junction material separately, to see how quantum shape effects influence their thermoelectric transport coefficients individually. By employing transmission formalism under the linear response regime \cite{dattabook1995,lundstrombook,norvec1}, the transport integral reads
\begin{align}
I_\alpha=&
	\int\left[\beta\left(\varepsilon-\mu\right)\right]^\alpha \beta f(\varepsilon)[1-f(\varepsilon)]{\cal T}(\varepsilon) d\varepsilon
	,
\end{align}
where $\beta=1/(k_B T)$ with Boltzmann constant $k_B$ and temperature $T$, $\alpha$ indicates the energy moment index, $\varepsilon$ is energy, $\mu$ is chemical potential, $f=1/\{\exp[\beta(\varepsilon-\mu)]+1\}$ is Fermi-Dirac distribution function and $\cal{T}(\varepsilon)$ is the total transmission function.

Calculation of total transmission can be done either by discretizing the region within the tight-binding approximation and the use of scattering matrix approach, or by using the Datta's number of modes formalism \cite{dattabook1995}. Under certain assumptions, both approaches give exactly the same result. We assume pure ballistic transport regime in this work, in order to maximize the effect of shape dependence \cite{bineker}. Due to their low effective mass, we consider n-type GaAs as our junction material. Since we explore thermoshape effect, we choose both pillars of the junction as the same type of material (n-type here). Effective mass of conduction electrons for GaAs nanostructures is $m^*=0.067 m_e$, where $m_e$ is the bare electron mass. Side lengths of square core and shell structures (Fig. 1) are chosen as 41nm and 64nm respectively so that de Broglie wavelengths (or Fermi wavelengths in weakly degenerate case) of electrons are on the order of domain sizes for the low temperature ranges (20K-50K) that are considered here to enhance the influence of geometry. While the mean de Broglie wavelength of particles is in the order of average sizes of the domain in transverse direction, it is much larger than the size of the surface roughness, which makes the roughness effect negligible or at least similar in both pillars of the junction so that surface roughness effects should be averaged out compared to the distinctness of shapes. 

The transport regime is chosen as fully ballistic so that the transport direction is smaller than phase coherence length, as well as both elastic and inelastic electron mean free paths. For GaAs, all these characteristic lengths can be found in Ref. \cite{bineker}. Longitudinal length can take any value as long as it is smaller than these characteristic lengths. Since at low temperatures, electron-phonon scattering is suppressed, we don't take it into account \cite{bineker}. Thermoelectric coefficients are calculated under zero net current condition, which means there shouldn't be applied bias on the system. Thus, we chose the leads and the scattering region to be translationally symmetric and considered zero bias. Due to these conditions, reflectance is zero and transmission coefficients for all transverse modes are exactly one. Using Datta's number of modes formalism, total transmission is the transmission of each mode multiplied by the number of modes (${\cal M}$) and since the transmission is one for every mode in purely ballistic regime, ${\cal T}(\varepsilon)={\cal M}(\varepsilon)=\sum_{\varepsilon_k}\Theta(\varepsilon-\varepsilon_k)$. $\varepsilon_k$ are energy eigenvalues from the solution of Schr\"odinger equation, which is implemented numerically by COMSOL Multiphysics software. Thus, under these assumptions, both tight-binding and number of modes approaches give the same results in our thermoshape junction \cite{dattabook1995,dattabook2005,datta1,datta2}. We compared the results with the ones of the tight-binding approximation with nearest-neighbor coupling. The numerical calculations are done on the quantum transport software KWANT \cite{kwant} and indeed we saw that tight-binding results converges to our results when lattice discretization parameter is chosen small enough.

By means of the transport integral, the electrical conductance, Seebeck coefficient and electronic thermal conductance are then written as \cite{dattabook1995,tebook2014b}
\begin{subequations}
\begin{align}
G=&
	\frac{2e^2}{h}I_0
	,
\\
S=&
	-\frac{k_B}{e}\frac{I_1}{I_0}
	,
	\\
\kappa_e=&
	\frac{2k_B^2}{h}T\left(I_2-\frac{I_1^2}{I_0}\right)
	,
\end{align}
\end{subequations}
where $e$ is electron charge, $h$ is Planck constant and 2 factors stand for spin degeneracy. Using the above thermoelectric transport coefficients one can easily find thermoelectric power factor by $GS^2$ relation. 

\begin{figure}[t]
\centering
\includegraphics[width=0.48\textwidth]{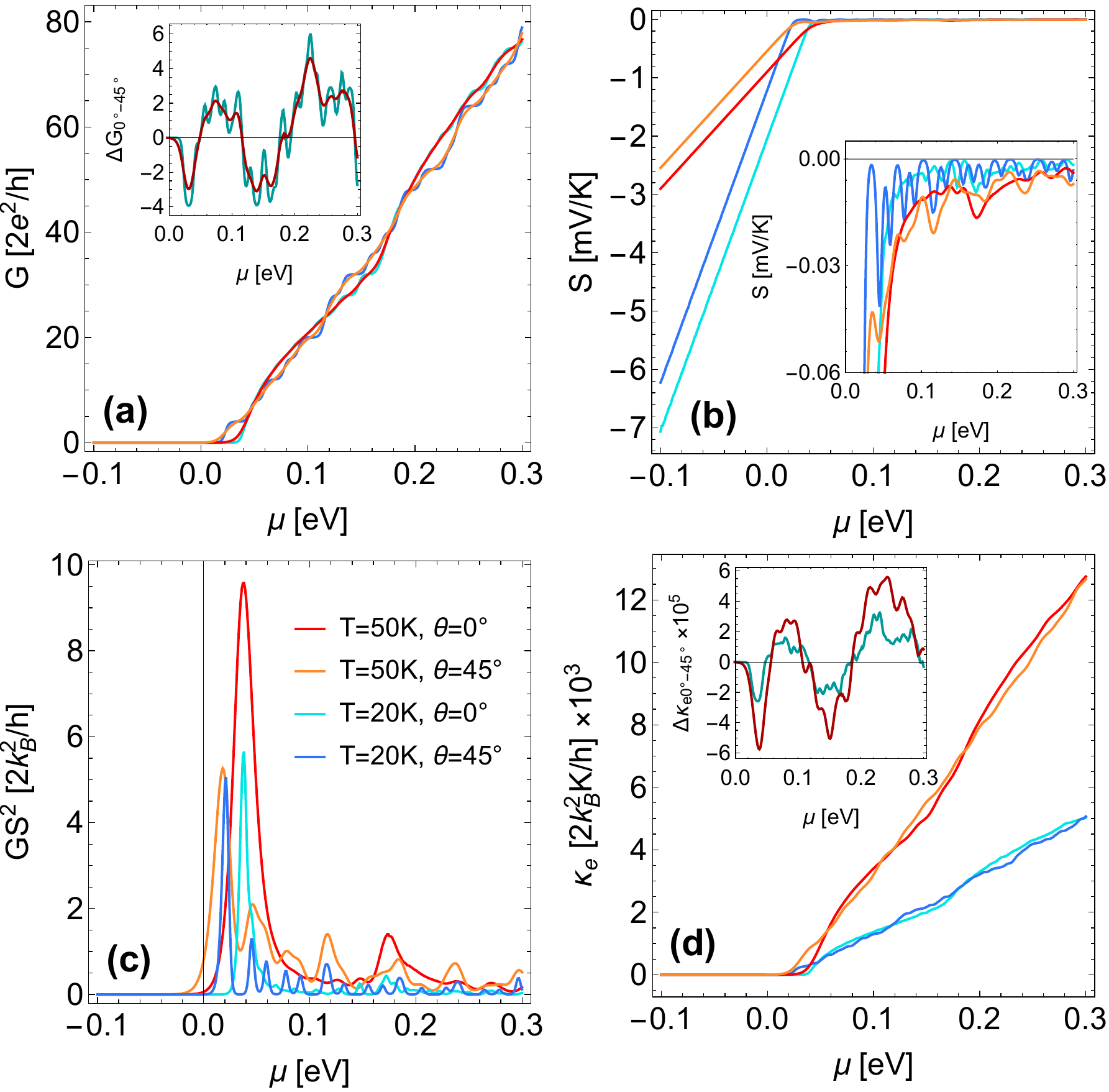}
\caption{(a) Electrical conductance, (b) Seebeck coefficient, (c) power factor and (d) electronic thermal conductance as function of the chemical potential for four different cases which are $T=20K, \theta=0^{\circ}$ (cyan), $T=20K, \theta=45^{\circ}$ (blue), $T=50K, \theta=0^{\circ}$ (red), $T=50K, \theta=45^{\circ}$ (orange). The legend given in (c) applies to all figures. Absolute differences of electrical and thermal conductances between two angular configurations are given as an inset in their respective figures for particular temperatures (teal color for $T=20$K, dark red for $T=50$K).}
\label{fig:pic2}
\end{figure}

In Fig. 2, variation of electrical conductance, Seebeck coefficient, power factor and electronic thermal conductance with respect to chemical potential are plotted. Chemical potential can be varied by controlling the doping concentration or the gate voltage. Cyan (red) and blue (orange) curves represent the values respectively for $\theta=0^{\circ}$ and $\theta=45^{\circ}$ configurations at $T=20$K ($T=50$K). The chemical potential ranges from -0.1 eV to 0.3 eV, which corresponds from non-degenerate case (between -0.1 eV and 0 eV) to weakly degenerate (between 0 eV and 0.3 eV) case. It is seen from Fig. 2 that, the results of $0^{\circ}$ and $45^{\circ}$ configurations are different than each other in all cases. Since the material and its sizes are the same, these differences are due to pure shape difference. In that sense, Fig. 2 represents the evidence of pure quantum shape dependence on thermoelectric transport coefficients of ballistic semiconductor nanostructures. To be more precise, shape difference causes a distinct change in energy spectrum, which changes the transmission as well the transport integral, along with the physical properties that depend on it. While conductance is directly related with both transmission and Fermi-Dirac distribution, energy moments in high order transport integrals (e.g. $I_1$ or $I_2$) cause more complicated behaviors in Seebeck coefficient and thermal conductance, Eqs. 2(a,b,c). 

Conductance quantization can be observed in some certain confined structures. Stepwise behavior of electrical conductance can be seen here clearly in $T=20K$ case, given by blue and cyan curves in Fig. 2a. For higher temperature case ($T=50K$), the steps become smoother as expected and stepwise behavior turns more into a collective 2D sub-band behavior. On the other hand, non-zero conductance appears at lower chemical potential for $\theta=45^{\circ}$ configuration (blue curve) in comparison with that of $\theta=0^{\circ}$ one (cyan). This shows that the effective confinement is lower in case of $\theta=45^{\circ}$ configuration. It is also seen that the stepwise behaviors of blue and cyan curves fluctuate around the orange and red curves respectively and the fluctuation amplitudes of cyan one are relatively smaller than those of the blue one in overall range. Absolute differences between conductances of two angular configurations are given as an inset in Fig. 2a. Teal and dark red colors represent 20K and 50K temperatures respectively. For lower temperature, variation due to quantum shape effects are larger than that of higher temperature one, suggesting that reducing temperature increases quantum shape effects as expected.

Variation of Seebeck coefficient with chemical potential for different configurations and temperatures are given in Fig. 2b. Two different characteristic behaviors can be clearly seen in different degeneracy regimes. From non-degenerate range to the beginning of quantum degeneracy (around zero chemical potential), Seebeck coefficients exhibit perfectly linear dependency on chemical potential. Difference in Seebeck coefficients due to quantum shape effects can also be clearly seen from the gaps in between the linear curves in the figure. For weakly degenerate chemical potential range, Seebeck coefficients have very small values near to zero and show oscillatory behavior, which can be seen from the inset in Fig. 2b. Oscillations are much frequent and strong for the lower temperature cases (cyan and blue curves) than those for higher temperature ones (red and orange curves). This shows that system becomes more sensitive to the variations in chemical potential in lower temperatures.

In Fig. 2c, thermoelectric power factor is given as a function of chemical potential. Due to near zero conductance in non-degenerate range (negative chemical potentials), power factor also gives zero. In all four cases, characteristic peaks of power factors can be seen around 0 eV and 0.1 eV ranges, which is an expected result. Since conductance for $\theta=45^{\circ}$ configurations start their first step earlier than $\theta=0^{\circ}$ ones (see Fig. 2a), power factor peaks of $\theta=45^{\circ}$ configurations appear closer to 0 eV than $\theta=0^{\circ}$ ones. This apparent distinction is also important for the utilization of thermoelectric power in thermoshape junctions. It can be seen that power factor for $\theta=0^{\circ}$ configurations are higher than those of $\theta=45^{\circ}$ ones, since the effective shape confinement is stronger at the former configuration. Note that the red peak is considerably higher than the orange one at the same temperature, since the magnitude of $S^2$ determines the magnitude of the peaks. Also, higher temperature peaks are less sharper than the lower temperature ones as expected.

Variation of electronic thermal conductance with chemical potential is also given in Fig. 2d. Higher temperature cases have larger thermal conductance than lower temperature ones as expected. This effect can also be seen in the inset of Fig. 2d where absolute differences between thermal conductances of two angular configurations are given. Although quantum shape effects are stronger at low temperatures, the absolute difference in thermal conductance is larger in $T=50$K case, because of the linear temperature factor in Eq. (2c), see inset figure of Fig. 2d. If one looks to relative differences, then lower temperature one ($T=20$K case) will have the higher amplitude.

From the results shown in Fig. 2 we see that even if the geometric size variables are the same, difference in shape alone modifies thermoelectric properties. In a thermoshape junction, on the other hand, two nanostructures with different shape configurations (one is prepared at $\theta=0^{\circ}$ and the other $\theta=45^{\circ}$) are made a junction. When temperature gradient applied to such system under zero external bias voltage, temperature difference becomes the only driving force for the charge current. Therefore, an electrochemical potential difference is induced as response to the driving force because of zero net current condition at steady-state. The net current inside each configuration of the thermoshape junction is given by
\begin{equation}
I_{net}=I_H-I_C=\frac{2e}{h}\int\left[f\left(\mu_H,T_H\right)-f\left(\mu_C,T_C\right)\right]\mathcal{T}(\varepsilon)d\varepsilon,
\end{equation}
where $\mathcal{T}(\varepsilon)$ is taken to be same for left and right going particles. Now, thermosize voltage can then be defined under zero net current conditions as the difference of electrochemical potentials at the hot end when they are connected at the cold end of the junction,
\begin{equation}
V_{TSh}(\mu_C,T_H,T_C,\theta_L,\theta_R)=\frac{1}{e}\left(\mu_H^{\theta_L}\big|_{I_{net}=0}-\mu_H^{\theta_R}\big|_{I_{net}=0}\right),
\end{equation}
where $\theta_L$ and $\theta_R$ are the angular configurations of left and right components of the junction shown in Fig. 1.

In Fig. 3, the chemical potential dependency of thermoshape voltage is shown in milivolt scale for 2K temperature difference at different cold side temperatures. The voltage is persistently give constant value from non-degenerate range until the 0.05 eV where weak degeneracy starts. Although the junction is made by the same material having the same sizes, the amount of the thermoshape voltage is still in several milivolts, just because of the shape difference.  After around 0.05 eV, thermoshape voltage reduces rapidly to near zero values and it has an oscillatory nature around zero voltage having both positive and negative values changeably, which can be seen in the inset of Fig. 3.

\begin{figure}[t]
\centering
\includegraphics[width=0.45\textwidth]{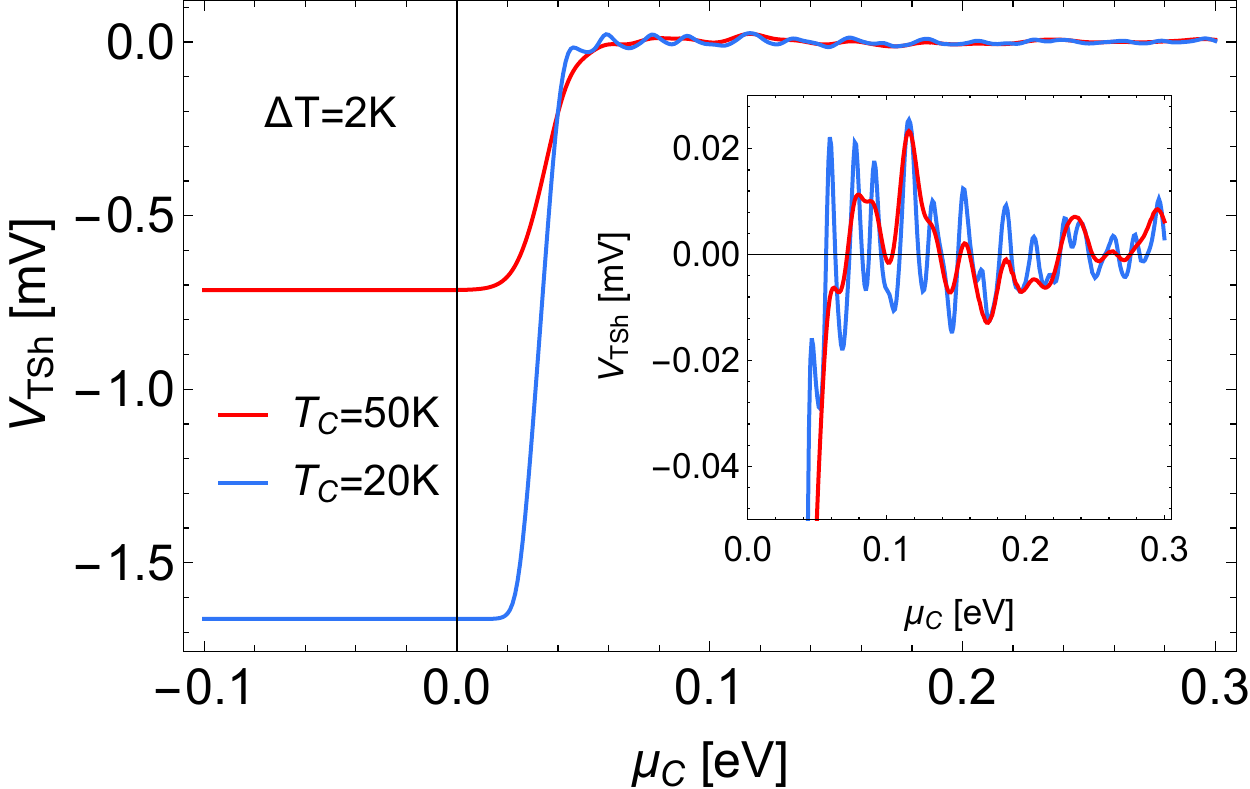}
\caption{Thermoshape voltage changes with the chemical potential of cold side for 20K and 50K cases. The thermoshape junction is made by $\theta_L=0^{\circ}$ and $\theta_R=45^{\circ}$ configurations of the same materials with the same sizes. Temperature differences between cold and hot sides are chosen as 2K. Inset shows the enlarged version of the oscillations in the higher chemical potential range.}
\label{fig:pic3}
\end{figure}

\begin{figure}[t]
\centering
\includegraphics[width=0.48\textwidth]{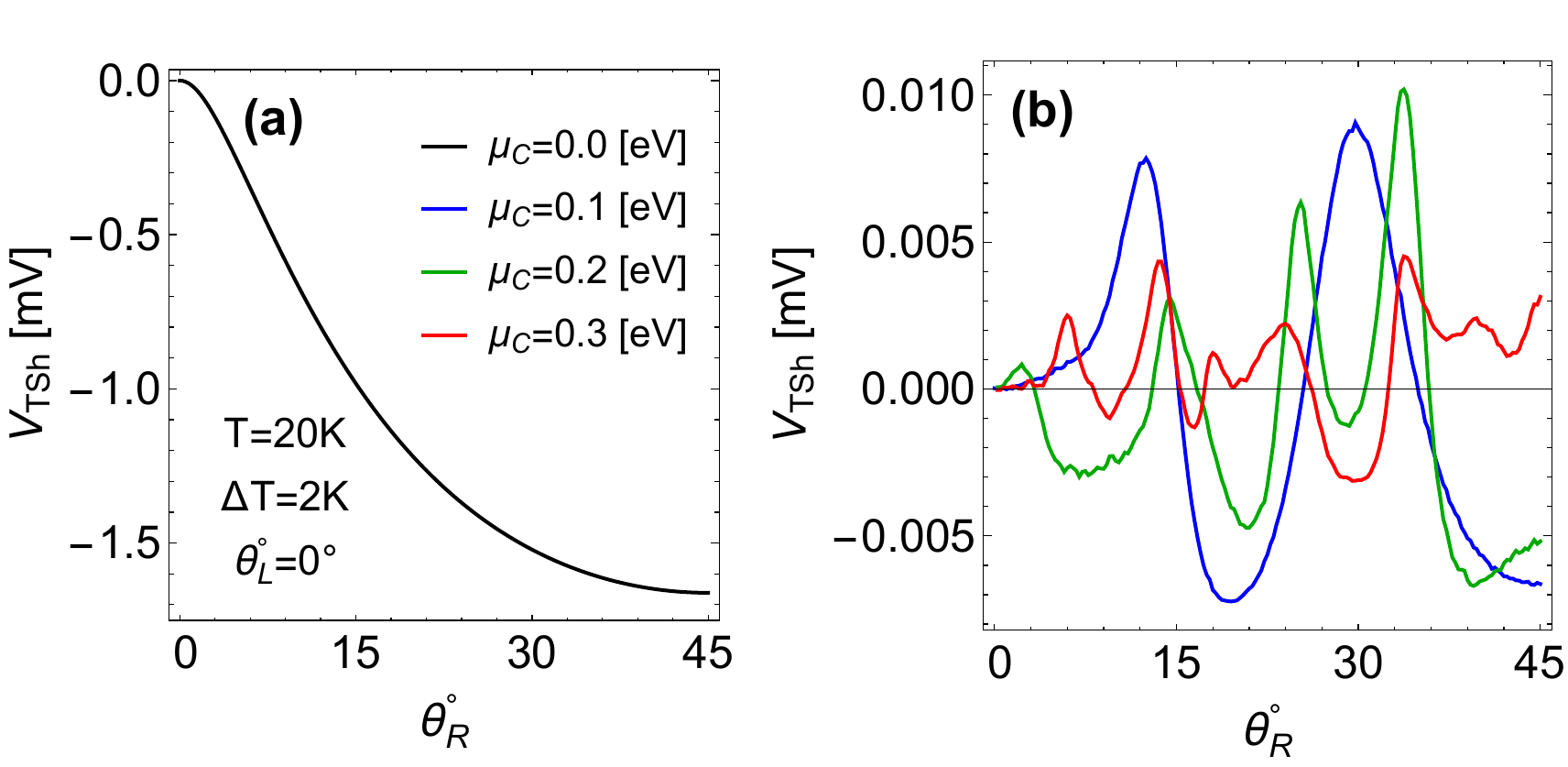}
\caption{Thermoshape voltage changes with the configuration angle of the core structure of the right component in the junction when the cold side chemical potential is (a) zero, (b) 0.1 eV, 0.2 eV and 0.3 eV, denoted by black, blue, green and red curves respectively at 20K and $\theta_L=0^{\circ}$. (a) 0 eV corresponds to the non-degenerate semiconductor regime and variation of thermoshape voltage with respect to configuration angle exhibits non-oscillatory behavior. Legends applies to all figures. (b) Higher chemical potential cases correspond to weakly degenerate semiconductor regime and thermoshape voltage exhibits oscillatory behaviors with a much less magnitude.}
\label{fig:pic4}
\end{figure}

For the calculations in this article, temperature difference of $\Delta T=2$K applied between both ends of the junction. The magnitude of the thermoshape voltage linearly increases with the increasing $\Delta T$. Conversely, thermoshape voltage magnitude is inversely proportional ($|V_{TSh}|\propto 1/T_C$) with the cold side temperature.

Rather than a thermoshape junction between $\theta=0^{\circ}$ and $\theta=45^{\circ}$ configurations, it may have been also arranged for intermediate configuration angles. The variation of thermoshape voltage by keeping the left component's core structure at $\theta=0^{\circ}$ configuration but changing the right configuration from $0^{\circ}$ to $45^{\circ}$ is given in Fig. 4 for 20K. At zero chemical potential, magnitude of thermoshape voltage gradually increase from zero to 1.5 mV range, Fig. 4a. This is because the higher the distinctness between the shapes (which is characterized by the change in $\theta$), the higher the thermoshape effects. In Fig. 4b, thermoshape voltage vs $\theta$ is given in weakly degenerate conditions for chemical potentials 0.1 eV, 0.2 eV and 0.3 eV, represented by blue, green and red curves respectively. At weakly degenerate regime, thermoshape voltage exhibits oscillatory behavior even with the variation of the configuration angle. This suggest that influence of shape on the physical properties of the electrons in confined systems is more complicated in degenerate regimes due to quantum shape dependent oscillations in chemical potential at constant number of particles. It should be noted that these oscillations are different than the ones controlled by chemical potential in Figure 2b. Oscillations seen in Fig. 4b are controlled by the angular configuration and they appear because of changes in effective confinement (a type of confinement controlled by change of shape). Physical explanation of changes in effective confinement can simply be done by considering overlapped quantum boundary layer concept and detailed explanations can be found in Ref. \cite{dsey}. Small cusps in Fig. 4 are due to negligible errors in numerical integrations.

Note that as long as the same metallic contacts are used in the thermoshape junction, their influence on the thermoshape voltage would be miniscule, because all kind of interface effects, including the one on the chemical potentials, in each pillar would almost be the same and should cancel their effects on thermoshape potential, since it is based on the differences in chemical potentials between two pillars.

\section{Conclusions}
In this article, we've proposed and presented the existence of an electric voltage induced by temperature and shape differences, so called thermoshape effect. The effect is similar to thermosize effect, though, rather than making junctions of materials with different sizes, we considered the junction of the same material with different shapes while keeping their sizes the same. Unlike thermosize effect, thermoshape effect is not a consequence of the direct confinement of the nanostructure, but the effective confinement due to overlaps of quantum boundary layers, see Ref. \cite{dsey} for more detail. For the considered material, sizes and temperatures; thermoshape voltage is on the order of several mV/K, which is in the same order with the electric voltage obtained from thermosize junctions. On the contrary to our first expectations, quantum shape effects are as strong as quantum size effects in thermoelectric-like junctions.

Although this study is focused on the proposition and examination of thermoshape effect for the first time, quantum shape effects not only give rise to thermoshape effect, they may also be used to enhance thermoelectric properties of the existing systems. For example instead of constructing a thermoshape junction, one may construct a thermoelectric junction with different materials and use quantum shape effects to make some enhancements on their thermoelectric performance. Graphene nanostructures are also very popular and convenient for thermoelectric applications at room temperature \cite{tegnr2015a}. As a room temperature candidate material, we will investigate the thermoshape effect in graphene thermoshape junctions as a future study. A possible enhancement of thermoelectric figure of merit in nanoscale thermoelectrics by quantum shape effects can be another important direction to be focused on.
\nocite{*}
\bibliography{tsheref}
\end{document}